\documentclass[twoside,a4paper,10pt]{article}
\usepackage{amsmath, amssymb, graphics, graphicx}
\usepackage{amscd}
\usepackage[dvips]{hyperref}
\usepackage{textcomp}
\begin{document}
\setlength{\topmargin}{-20pt}
\setlength{\evensidemargin}{20pt}
\setlength{\oddsidemargin}{20pt}
\setlength{\textheight}{9.2in}
\title{A model of returns for the post-credit-crunch reality: Hybrid Brownian motion with price feedback}
\author{William T. Shaw\thanks{Corresponding author: Department of Mathematics
King's College, The Strand,
London WC2R 2LS, England; E-mail: william.shaw@kcl.ac.uk}}

\maketitle
\begin{abstract}
The market events of 2007-2009 have reinvigorated the search for realistic return models that capture greater likelihoods of extreme movements. In this paper we model the medium-term log-return dynamics in a market with both fundamental {\it and technical} traders. This is based on a Poisson trade arrival model with variable size orders. With simplifications we are led to a {\it hybrid} SDE mixing {\it both arithmetic and geometric} Brownian motions, whose solution is given by a class of integrals of exponentials of one Brownian motion against another, in forms considered by Yor and collaborators.  The reduction of the hybrid SDE to a single Brownian motion leads to an SDE of the form considered by Nagahara, which is a type of ``Pearson diffusion'', or equivalently a hyperbolic OU SDE.  Various dynamics and equilibria are possible depending on the balance of trades. Under mean-reverting circumstances we arrive naturally at an equilibrium fat-tailed return distribution with a Student or Pearson Type IV form. Under less restrictive assumptions richer dynamics are possible, including bimodal structures. The phenomenon of variance explosion is identified that gives rise to much larger price movements that might have a priori been expected, so that ``$25\sigma$'' events are significantly more probable. We exhibit simple example solutions of the Fokker-Planck equation that shows how such variance explosion can hide beneath a standard Gaussian facade. These are elementary members of an extended class of distributions with a rich and varied structure, capable of describing a wide range of market behaviours. Several approaches to the density function are possible, and an example of the computation of a hyperbolic VaR is given. The model also suggests generalizations of the Bougerol identity.
\end{abstract}

\pagestyle{myheadings}
\thispagestyle{plain}
\markboth{W.T. Shaw}{Price feedback and hybrid BM: a return model for the PCC reality}

\newpage
\section{Introduction}

{\center{\it `Technical analysis is an anathema to the academic world.'}\\
\medskip
Burton Malkiel, in ``A Random Walk Down Wall Street''.}
\medskip

The fat-tailed and non-normal distribution of share-price returns has been well known for decades. Mandelbrot \cite{mandel} and Fama \cite{fama} noted the excess kurtotic nature of equity returns in the 1960s. The need for non-(log)Gaussian has been widely recognized by numerous authors, and the events of 2007-2009 have made it very clear that extreme share-price movements are much more likely than in a simple log-normal model associated with geometric Brownian motion. Despite the fat-tailed behaviour being known for 40 years many market participants cling to the Gaussian or near Gaussian picture, and attempt to explain away the large movements while remaining within that picture. The CFO of Goldmans, in a {\it Financial Times} article, famously attempted to excuse the implosion of Goldman's hedge funds with the comment\footnote{David Viniar, Goldman's chief financial officer, as quoted by Peter Larsen, Financial Times August 13 2007.}

{\it ``We were seeing things that were 25-standard deviation moves, several days in a row''}.

These were {\it not} events that can be written off as un-modellable due to their extreme unlikelihood - people were just thinking of the wrong, Gaussian, distribution based on an unrealistic world-view, while simultaneously discounting the possibility of both major shifts in the mean and sudden increases in variance. For example, the likelihood of a single $25\sigma$ or worse event, without allowing for a major shift in the distributional mean or variance, is 
\begin{itemize}
\item about $6 \times 10^{-138}$ in a Gaussian picture;
\item about $4 \times 10^{-6}$ in a Student-t picture (four degrees of freedom, as estimated for global indices in \cite{platen}); 
\item about $1/625$ for a toss of a very unfair coin where the probability of a head is $1/625$.\footnote{This example is based, with thanks, on an anonymous observation at \url{http://worldbeta.blogspot.com/2007/08/really-with-seth-and-amy-part-ii.html}}
\end{itemize} 

The estimation of the observed return distribution is a matter for proper statistical analysis. The emergence of some form of fat-tailed or indeed power-law decay in the tails is a robust feature on various time-scales. The leading contenders for the best fit are often a variance-gamma (VG) distribution or a Student t, with some evidence for normal-inverse Gaussian. A useful approach is to do a maximum likelihood estimate within the generalized hyperbolic family.  See, for example, the recent study by Taylor {\it et al} on South African index data \cite{taylor},  Fergusson and Platen \cite{platen}, and references therein. Extreme events are much more common in such models. The work of \cite{platen} found a Student with about four degrees of freedom ($\nu=4$), and the work of \cite{taylor} ranked a Student as second most likely with higher values of $\nu$. General non-integer low values of $\nu$ may well be of interest in financial analysis for short time-scales. Work cited in \cite{highfreq} suggests that very short term returns exhibit power law decay in the PDF.  The values of $\nu$ reported in \cite{highfreq} take values in the range $2$ to $6$. So this leads us to consider not only small integer values of $\nu: 2 \leq \nu \leq 6$ but  also non-integer $\nu$. In normal statistical estimation use the $\nu$ parameter is usually assumed to be integer as it relates to sample size, but there is no underlying mathematical reason for this restriction.

The huge losses in the markets in the fallout of the credit crunch were due to several factors, which include not only the very real loss in value of financial institutions who might, for example have substantial CDS exposure or exposure to genuine economic downturn, but also to the fear of traders panicking to offload falling assets. 

But long before the recent panic, participants in financial markets traded according to diverse strategies. Two common approaches are the so-called ``fundamental'' and ``technical'' trading strategies. Traders in the first group are largely interested in measure of value and traders in the second group are interested more in the dynamics of the price (and volume) history. In this note the impact of having these two types of trades is analyzed employing a simple trade arrival model with Poisson characteristics, with variable size orders. In general we obtain a large class of price evolution models. 

The approach presented here has conceptual links with other approaches. In particular, Brody, Hughston and Macrina (`BHM') \cite{bhm} have introduced the notion of {\it information-based asset pricing}, paying rather more attention to the filtration component of processes than has hitherto been developed. In the model presented here we package information received since the start of a trading period into the price history since the trading start, on the basis that market participants who believe in technical trading are at work.\footnote{This of course makes {\it no} judgement as to the wisdom, mathematically, statistically or otherwise, of technical trading, but does acknowledge its presence. In fact all we really need is the presence of {\it price-sensitive} orders. In reality the weight of technical trading is growing, with complex program trades taking up an increasing fraction of the market. Even fundamental traders will demand efficient execution and hence acquire a degree of price-sensitivity. The model developed here is a simple, linear and Markov representation of this market complexity.}  Our linearized model results in an SDE that is, at least at the conceptual level, a linear price-information correction to standard Brownian motion of log-returns. It remains to be seen whether this notion can be given a rigorous mathematical basis within the BHM framework. 

We consider the dynamics on a medium time scale which may be considered as assessing the distribution of returns on a daily basis as the aggregate of many intra-day trades. Under certain simplifying assumptions, corresponding to linearization of price impact functions, linearization of technical trading criteria, and a ``many trade'' limit, we are lead to an SDE that is a {\it hybrid of arithmetic and geometric Brownian motion}. Various dynamics and equilibria arise from this SDE  depending on the balance of trades. Under mean-reverting circumstances we are lead naturally to an equilibrium fat-tailed return distribution with a Student $t$ or skew-Student form, with the latter defined within the framework of ``Pearson diffusions'' defined by Forman and S{\o}rensen \cite{forman}. The construction of the standard Student distribution from an SDE has also been recently considered by Steinbrecher and Shaw \cite{quantile}. The model is capable of still richer dynamics, and in more generality leads to a hyperbolic extension of the OU process. In general form a non-linear diffusion with or without jumps is possible. One effect of note is the simultaneous explosion in the variance coupled to the emergence of a non-Gaussian distribution, and such distributions hiding their character under a Gaussian mask. When there is significant momentum bimodal densities may arise.

This note is not to first to propose a model with Student equilibria. Indeed the work by Nagahara \cite{nagahara}, that postulated a certain Pearson diffusion, pre-dates our analysis significantly, and the associated density function was given by Wong \cite{wong} many years ago (it is Wong's ``Type 2E''). The work by Carmona, Petit and Yor  \cite{carmonayor}, set in the context of L\'{e}vy processes, contains closely related structures. However, the full financial market and risk significance of this model, and its potential justification in terms of {\it price feedback via technical or similar trading}, does not appear to have significant penetration in the financial mathematics community, and the technology associated with managing the resulting densities is not developed.  In this current note we provide: 
\begin{enumerate}
\item a sketch of a microstructure justification;
\item the connection with the theory of integrals of exponentials of Brownian motion;
\item an alternative density representation, and some simplified forms of the density;
\item an analysis of market states and pathological cases;
\item an initial approach to the VaR analysis.
\end{enumerate}
In this way we hope the reader will be persuaded of the rationale, underlying simplicity and practical applicability of this model, which produces fat-tailed behavior consistent with market behaviour in a simple manner. There is of course a jump-extended version of the model incorporating technical trading. Here we do not allow for individual ``large'' trades. Research into more tractable forms of Wong's distribution \cite{wong,nagahara} is also needed. We will offer a simple single-term formula for the transform here. The theory of integrals of exponentials of Brownian motion, and of L\'{e}vy processes, has attracted much study over the years and is of considerable interest from the point of view of both pure analysis and practical applications, initially to the Asian option. The reader is referred to the articles by Dufresne \cite{dufresne, dufresne01} and to  the classic collection of papers by M. Yor \cite{yorbook} and the references therein for the detailed mathematical background\footnote{The author is indebted to Prof M Yor for pointing out many of the relevant papers on related topics - any omissions that remain are due to me.}. Several properties of integrals of GBM are also reviewed in \cite{dmmy}. A more recent survey of some key ideas is given by Matsumoto and Yor \cite{matsumoto}, see also \cite{my2003}. For the PDE enthusiast we refer the reader to Dewynne and Shaw \cite{ejam}, where an unusually quick derivation of the Asian Call price is given together with an volatility series, cf. Zhang \cite{zhangb}.  There are many other papers on approximations, bounds, trading, discrete averaging but the key papers above and references therein are most relevant to the exact representation in terms of integrals of exponentials of Brownian motion. For practitioner applications many discrete and continuously averaged cases of Asian options may be computed by the approach of Vecer \cite{vecer}.

In general the probability density functions arising from the model are complicated. The direct Laplace transform approach presented here offers an alternative representation to that given by Wong \cite{wong}, and has the advantage that several special cases are closed-form computable for both mean-reverting and momentum-dominated states with integer drift. This includes both some cases (mean-reverting) identified by Wong, as well as simple momentum-dominated cases intimately related to the {\it Bougerol identity} in both the original form, \cite{bougerol}, and the beautiful bi-modal variation found by Alili, Dufresne and Yor \cite{revistapaper}. It is hoped that the representations given here may lead to further results along these lines. 

\subsection{Analogies from the physical and biological world}
We will work with an SDE that is a {\it hybrid} of arithmetical and geometric Brownian motions. This SDE has many interesting features, and has an interesting history in the physics literature\footnote{I am grateful for Dr. G. Steinbrecher for these insights.}. A very closely related analysis has been given quite recently in the plasma physics literature \cite{gsplasma}. However, the history of hybrids of both arithmetical and geometric Brownian motions dates back to the 1979 paper by Schenzle and Brand \cite{physreva}. 
Also, discrete models in the form of a linear Langevin equation:
\begin{equation}
x(t+1) = b(t) x(t) + f(t)\ ,
\end{equation}
where $b$ and $f$ are both random, were considered in a thread of research originating in 1997 with Takayasu {\it et al}, who established conditions for the realization of power law behaviour \cite{prl}. 

On the biological side, May {\it et al} \cite{maynature} have observed the ``common ground in analysing financial systems and ecosystems'', and identify feedback mechanisms as one mechanism for causing catastrophic changes in the overall state of a system, with hidden linkages being another cause. Here we are focusing on the mechanism of price feedback within the financial context, while adapting the familiar financial mathematics framework of stochastic differential equations to cope with it. The importance of feedback has also been emphasized by Kambhu {\it et al} \cite{kambhu}, in a major survey encouraging the financial community to draw on the lessons from ecology, atmospheric science and other complex systems. It is hoped that this current paper will contribute to that discussion.
 
\subsection{Plan of article}

The plan of this work is as follows. In Section two the underlying dynamics of a market containing both fundamental and technical traders is considered. In Section three the model is simplified and linearized, and approximated to a tractable model. In general a model with a combination of jumps and pure Brownian motion is possible, but we focus on the non-jump case for further analysis. In Section four the possibility of a Student t equilibrium is identified. In Section five the notion of a ``hyperbolic OU'' process is established, and in Section six the possibility of Pearson Type IV is considered. The fully dynamical case is analyzed in Section seven. In particular the phenomenon of ''variance explosion'' is identified. We also demonstrate the nature of the full distribution by reference to an explicitly solvable special case, linked to Bougerol's identity. In Section 8 we analyze the general case, including examples of both momentum-dominated and mean-reversion-dominated markets. This section also explores other closed-form distributional solutions and gives a Laplace transform solution for a large class of cases, revealing the link to the Legendre equation well known in physics,  and to the Bougerol identity and its bimodal generalization. Section nine offers a preliminary market classification based on this approach. Section ten explores the VaR implications and Section eleven offers conclusions and speculations. 

\section{Two types of market participant}
This section is intended to be a sketch of a series of links from basic trading ideas through to a stochastic differential equation, and is intended in that spirit only. There are aspects where a more rigorous mathematical treatment will ultimately be needed, such as the details of the SDE limiting process. However, we think it is important to provide such a sketch. The later analysis could proceed basely solely on the postulation of the resulting hybrid SDE, but it is intended that this paper provide an end-to-end discussion, starting with ideas about trades and ending with a density function.

Consider a market containing an asset with share price $S_t$ at time $t$. We consider a trading period  $t \in [0, T]$. The log-return, $x_t$, on the asset is given by
\begin{equation}
x_t =\log\left(\frac{S_t}{S_0}\right)\ ,
\end{equation}
and we shall be concerned with time intervals over which $S_t \sim S_0(1+x_t)$ so that no distinction is made between log and linear returns. 

{\it Our approach is to consider that the agents trading in this asset have two different types of motivation.}

One set of agents comprise those acting independently of the current value of $x_t$. These will include, but will not necessarily be limited, to those trading on fundamentals, such as the dividend or earnings yield based on the price $S_0$. Other such traders may be engaged in a portfolio rebalance, or re-hedging a derivative position.  

The second set of agents trade on the basis of the value of $x_t$. These are our technical traders relative to the time period under consideration. Technical trading is based on a number of different algorithms, which may be classified according to a ``Rumsfeld'' scheme: price momentum, mean reversion (known knowns) through unknown unknowns such as black box unpublished models being run by covert trading operations. We do not know the totality of such agents but we do know that they care about $x_t$, or possibly about the history $\{x_s|0\leq s \leq t\}$. In this paper I shall simplify and consider traditional momentum and mean-reversion traders who act on the value of $x_t$. There may also be longer term technical traders focusing on the ratio of $S_0$ to some much older price, who happen to be at work in the given time interval, but for our purposes they will not count as technical as they are not reacting specifically to $x_t$. There may be other participants who would not wish to be considered as technical traders, e.g those carefully achieving a position by a sequence of trades, but to the extent that their willingness to trade depends on $x_t$ they are technical. Neither shall we enter the debate as to whether technical trading makes any sense, whether price increments are independent or whether there are material serial correlations etc. etc. {\it It will suffice that people who believe in technical trading {\bf are} trading}. In this sense Malkiel's famous objection is irrelevant.

\subsection{Fundamental ``buy'' orders }
Let us now consider a time interval $(t, t+\Delta t) \subset [0,T]$ with $\Delta t <<  T$, and that orders may be effected in lots of size $L$.  {\it We consider first {\bf buy} orders based on {\bf fundamental} trading.} 

Let $Y$ be the integer-valued random variable denoting the number of such trades arriving in time $\Delta t$, and let $N_i, i=1,\dots Y$, be the integer-valued random variable denoting the number of lots in each buy order.  The number of shares, $M_B$, in the total collection of buy orders is $M=L \times Z$, where the random variable $Z$ is given by
\begin{equation}
Z =  \sum_{i=0}^Y N_i\ .
\end{equation}
We assume that the number of trades is independent of the size of each trade, and that the trade sizes are independent and identically distributed\footnote{These {\it are} assumptions and ones that might reasonably be questioned. One might consider, for example, that in a significant market downturn, there is correlation between having larger trades and having more trades. But even in this case we can imagine many small investors selling as well. Our goal is to get a model of price feedback, and the elegant compositional relationship for PGFs that these assumptions enable allow us to proceed more easily, if not in complete generality.}. Then the probability generating function (PGF) $f_Z(s)$ of $Z$ is related to the PGFs of $Y$ and $N$ by
\begin{equation}
f_Z(s) = f_Y(f_N(s))\ ,
\end{equation}
from which elementary PGF theory tells us that
\begin{equation}
E[Z] = E[Y] E[N] = E[Y] \overline{n}\ ,
\end{equation}
where $E[N] = \overline{n}$, and
\begin{equation}
Var[Z] = Var[Y] \overline{n}^2 + E[Y] Var[N] \ .
\end{equation}
We shall now assume that the buy variable $Y$ either follows a Poisson process with arrival rate $\lambda_B$, or, more loosely, that the process is sufficiently Poisson-like that we can write
\begin{equation}
E[Y] = Var[Y] = \lambda_B \Delta t\ .
\end{equation}
In fact what we really need is only that this last equation holds. $Y$ following a Poisson process is sufficient but not necessary. So then we have
\begin{equation}
E[Z] =  \lambda_B \Delta t E[N] =  \lambda_B \Delta t \overline{n}\ ,
\end{equation}
\begin{equation}
Var[Z] =  \lambda_B \Delta t (E[N]^2 + Var[N]) =  \lambda_B \Delta t (\overline{n}^2 + Var[N]) = \lambda_B \Delta t E[N^2]\ .
\end{equation}
It follows that
\begin{equation}
E[M_B] = L \lambda_B \Delta t \overline{n}\ ,
\end{equation}
and that the standard deviation of $M_B$, $sd(M_B)$, is
\begin{equation}
sd(M_B) = L \sqrt{\lambda_B \Delta t E[N^2]}\ .
\end{equation}

\subsection{Fundamental ``sell'' orders}
This proceeds in the same way. With similar assumptions, including that the variable $N$ is similarly distributed, we end up with mean of the number of sales in the sell orders as
\begin{equation}
E[M_S] = L \lambda_S \Delta t \overline{n}\ ,
\end{equation}
and that the standard deviation of $M_S$, $sd(M_S)$, is
\begin{equation}
sd(M_S) = L \sqrt{\lambda_S \Delta t E[N^2]}\ .
\end{equation}

\subsection{Aggregation of fundamental trades}
We define
\begin{equation}
M_F = M_B-M_S
\end{equation}
as the net buy volume. The random variable $M_F$ has mean
\begin{equation}
E[M_F]=L(\lambda_B-\lambda_S)\Delta t \overline{n}\ .
\end{equation}
and its variance depends on the correlation between arrival rates of buy and sell trades. If we assume that fundamental buyers and sellers are acting on rather different motivations, which seems reasonable, then we might assume independence and hence that
\begin{equation}
Var[M_F]=L^2(\lambda_B+\lambda_S)\Delta t  E[N^2]\ .
\end{equation}

\subsection{The technical traders}
We consider now a second group of traders who only trade in response to a return created within the period under consideration. We shall model these trades in the same way as for the fundamental case, except now the trade arrival rates, instead of $\lambda_{B,F}$ are now $\mu_B(x)$ and $\mu_S(x)$ where the buy and sell $\mu_i$ have the property that $\mu_i(0)=0$ - we assume that in the absence of a price movement there are no technical trades. Treating the variation in trade size in an identical fashion, we have additional net buying pressure $M_T$ which is a random variable with
\begin{equation}
E[M_T]=L(\mu_B(x)-\mu_S(x))\Delta t \overline{n}\ ,
\end{equation}
and a variance which again depends on the degree of correlation between the buy and sell actions. In this case we shall assume that the buy and sell actions are perfectly correlated, which makes sense for example if a trader is pursuing a mean-reverting strategy:
\begin{equation}
Var[M_T]=L^2(\mu_B(x)-\mu_S(x))\Delta t  E[N^2]
\end{equation}
It is natural to assume that the technical traders are operating independently from the fundamental traders and this will be done.

\subsection{The return impact function}
We introduce a log-return impact function ${\cal I}(q)$ that is a function with ${\cal I}(0)=0$ such that the aggregate buy and sell orders of both types create a log-return impact of the form
\begin{equation}
\Delta x = {\cal I}(M_F+M_T)\ .
\end{equation}

\subsection{Summary of discrete model}
In the time interval $\Delta t$ the return changes by
\begin{equation}
\Delta x = {\cal I}(M_F+M_T)\ ,
\end{equation}
where ${\cal I}(q)$ is the return impact of a net order to buy q shares. The variable $M_F$ is a random variable with mean
\begin{equation}
E[M_F]=L(\lambda_B-\lambda_S)\Delta t \overline{n}
\end{equation}
and variance
\begin{equation}
Var[M_F]=L^2(\lambda_B+\lambda_S)\Delta t  E[N^2]\ .
\end{equation}
The variable $M_T$ is independent of $M_F$, and has mean
\begin{equation}
E[M_T]=L(\mu_B(x)-\mu_S(x))\Delta t \overline{n}
\end{equation}
and variance
\begin{equation}
Var[M_T]=L^2(\mu_B(x)-\mu_S(x))\Delta t  E[N^2]\ .
\end{equation}

\section{Linearization and the SDE}
In order to proceed to a continuum representation we shall now make some simplifying assumptions. These are
\begin{enumerate}
\item Linearization of the return impact;
\item Linearization of the $\mu$ functions;
\item a Brownian motion model.
\end{enumerate}
We now consider each of these assumptions.
\subsection{Impact linearization}
For the return impact function all we know for sure is that no orders mean no price impact, i.e. ${\cal I}(0)=0$. In general ${\cal I}(q)$ may be a very complicated function. We think it should not decrease as $q$ increases. The structure of the order book may give it a staircase character\footnote{Note we always work based on mid-prices so the central step is taken as absent.} However, we shall assume that on certain scales we may linearize it - a staircase may look like a smooth line viewed from a ``long way'' away - in order to capture the gross feature of the model. So for some unknown constant $\omega$ we write
\begin{equation}
\Delta x = \omega \times (M_F+M_T)\ .
\end{equation}
The notion of a linear model of price impact was employed by Almgren and Chriss \cite{alm} as a component of both temporary and permanent price impact. 
It should be appreciated that this is very much in the same spirit as the assumptions made in the Bakstein-Howison model \cite{bakhow}, where the essential features of the order book are reduced to a spread and liquidity parameter pair, where liquidity is the reciprocal of the slope of the order book. This idea is a powerful one which may also be used to analyze derivatives, as considered e.g. by Mitton \cite{mitton}. Thus liquidity is an implicit part of the model via the linearized price-impact.

\subsection{Technical trade linearization}
The appearance of trades depending on price movements within the time period under consideration will also have a complicated dependence. The existence of limit orders at various price thresholds will also create a staircase effect. We shall linearize this in the same way as the price impact, and make the replacement:
\begin{equation}
\mu_B(x)-\mu_S(x) \rightarrow -\mu \times x\ ,
\end{equation}
where $\mu$ is an effective constant parameter that captures the gross slope of the function\footnote{In both this case and the treatment of price impact we are {\it not} making differentiability assumptions and using a power series - the idea in both cases is that a complex staircase may be grossly idealized as a sloping plane.} We put in a minus sign due to the nature of limit orders coming into play to act against the direction of price movement. One would expect the effect of $\mu$ to be negative on balance due to the effects of profit taking as well, {\it unless the system is being overwhelmed by momentum trades}. The appearance of the latter effect will be analyzed in detail later in this paper.

\subsection{Process approximation}
The trade arrival model is at this stage purely of Poisson type with a variable trade size. There are a number of ways in which this might be managed, depending on the details and frequency of trade arrivals. If there are a sparse number of large trades, this will be best viewed as a pure jump process. We might also have a situation where a large number of trades of moderate size together with sporadic large ones. This will generate in effect a jump diffusion. In the following simplification we assume that there are a large number of  trades of moderate size so that we do not consider the jumps. The other cases will be investigated elsewhere. It is also possible, indeed likely, that the net price impacts of the two types of trade and their associated volatilities can be time dependent - this model does not rule out stochastic volatility at all. However, in what follows we shall confine attention to constant parameters and the pure-diffusion view - even this subset of possibilities will demonstrate a rich structure through the emergence of a {\it hybrid arithmetic-geometric} stochastic process. 

In this context we now approximate the Poisson-type trade(s) arrival model(s) by independent Brownian motions centred on the mean arrival rate. We are lead finally to a discrete-time stochastic evolution equation for the return:
\begin{equation}
\Delta x = \omega L \biggl[\overline{n}[(\lambda_B-\lambda_S)-\mu x]\Delta t + s_1 \Delta W_1  + s_2 x \Delta W_2  \biggr]\ ,
\end{equation}
where
\begin{equation}
s_1 = \sqrt{(\lambda_B+\lambda_S)E[N^2]}\ ,\ \ \ s_2 = \sqrt{\mu E[N^2]}\ .
\end{equation}
We now take the continuum limit and finally arrive at the SDE

\begin{equation}
dX_t = (\mu_1-\mu_2 X_t) dt + \sigma_1 dW_{1t}  + \sigma_2 X_t dW_{2t}\ ,
\end{equation}
where
\begin{equation}
\begin{split}
\mu_1 &= \alpha \overline{n}(\lambda_B-\lambda_S)\ ,\\
\mu_2 &= \alpha \mu \overline{n}\ ,\\
\sigma_1 &= \alpha \sqrt{(\lambda_B+\lambda_S)E[N^2]}\ ,\\
\sigma_2 &= \alpha \sqrt{\mu E[N^2]}\ ,\\
\end{split}
\end{equation}
and $\alpha = L \omega$ is the return impact of trading one lot of shares. The SDE given above is the basic description where we show the explicit contribution separately of the fundamental and technical trades. We can of course reduce it to an SDE with a single noise term as follows. If $\rho$ is the correlation between the two Brownian motions, then we can write the SDE as
\begin{equation}
dX_t = (\mu_1-\mu_2 X_t)dt + \sqrt{\sigma_1^2 + X_t^2 \sigma_2^2 + 2 \rho \sigma_1 X_t \sigma_2^2}\  dW_t \ .
\end{equation}
This is one of the class of ``Pearson diffusions'' considered by Forman and S{\o}rensen \cite{forman}. The first {\it detailed} application to financial modelling that the author is aware of is the work by Nagahara \cite{nagahara}. It has a notable special case that we now consider.

\section{The Student equilibrium model}
A particular case of interest is obtained by considering $\mu_1 = 0 = \rho$, so that we obtain the SDE
\begin{equation}
dX_t = -\mu_2 X_t dt + \sqrt{\sigma_1^2 + X_t^2 \sigma_2^2}\  dW_t \ .
\end{equation}
In the equilibrium situation, the quantile ODE \cite{quantile} associated with this SDE reduces to
\begin{equation}
\frac{\partial^2 Q}{\partial u^2} \biggl(\frac{\partial Q}{\partial u}  \biggr)^{-2} = \frac{2(\sigma_2^2+\mu_2)Q}{\sigma_1^2 + \sigma_2^2 Q^2}\ .
\end{equation}
Bearing in mind the results of \cite{quantile} we see that we have a quantile function for a Student distribution with 
\begin{equation}
Q = \frac{\sigma_1}{\sqrt{\sigma_2^2+2\mu_2}}w(u)\ ,
\end{equation}
where $w(u)$ is the standard Student quantile with degrees of freedom
\begin{equation}
\nu = 1+2\frac{\mu_2}{\sigma_2^2}\  .
\end{equation}
So it is clear that we need $\mu_2>0$ for this to be a Student distribution. This of course corresponds to the requirement that the underlying SDE mean-revert to the origin, and this mean-reversion condition in turn allows an equilibrium to establish.  This equilibrium origination of the standard Student distribution arises naturally in plasma physics \cite{gsplasma}.  The faster the mean-reversion rate is compared to the multiplicative volatility, the closer the system is to the normally distributed limit. The Student distribution also arises naturally in the modelling of asset returns \cite{platen,shawjcf06,taylor}. 

In more generality we obtain a dynamic hyperbolic generalization of an OU process. This has been argued by Forman and S{\o}rensen to be, in equilibrium, a natural candidate for a skew-Student model. In complete generality a still richer class of diffusions with or without jumps is possible. 

\section{The ``hyperbolic O-U'' SDE}
While we do not have a characterization of the full time-dependent aspects, some insight can be gained by reducing one form of the hybrid SDE to standard form. First we do some scalings to standardize Eqn.(31). We let
\begin{equation}
X_t = \frac{\sigma_1}{\sigma_2 \sqrt{\nu}}Y_t\ ,\ \ \Sigma_0 = \sigma_2 \sqrt{\nu}\ .
\end{equation}
Then the SDE for $Y_t$ is
\begin{equation}
dY_t = -\frac{\Sigma_0^2}{2}\biggl(1 - \frac{1}{\nu}\biggr)Y_t dt + \Sigma_0\sqrt{1+\frac{Y_t^2}{\nu}}dW_t\ ,
\end{equation}
and this is a simple two-parameter form of the problem. We can re-cast this by setting
\begin{equation}
Y_t = \sqrt{\nu} \sinh(Z_t)\ ,
\end{equation}
and this leads to
\begin{equation}
dZ_t = -\frac{\Sigma_0^2}{2}\tanh(Z_t)dt + \frac{\Sigma_0}{\sqrt{\nu}}dW_t\ ,
\end{equation}
or equivalently
\begin{equation}
dZ_t = -\frac{\sigma_2^2}{2} \nu \tanh(Z_t)dt + \sigma_2dW_t \ .
\end{equation}
With a time-scaling we have the non-dimensional form:
\begin{equation}
dZ_{\tau} = -\frac{1}{2} \nu \tanh(Z_t)d\tau + dW_{\tau} \ .
\end{equation}
The original variable $X_t$ is then given in terms of $Z_t$ as simply:
\begin{equation}
X_t = \frac{\sigma_1}{\sigma_2} \sinh(Z_t)\ .
\end{equation}
One can explore the full Fokker-Planck equation based on either of equations (36) or (39).  Equation (39) reveals the essential nature of the process: for small $Z_t$ and small times, the process is essentially OU in character. But the mean-reversion in the tails levels off and becomes much weaker. 

\section{Pearson type IV: skew-Student equilibria}
We now turn to the more general case, where the SDE is
\begin{equation}
dX_t = (\mu_1-\mu_2 X_t)dt + \sqrt{\sigma_1^2 + X_t^2 \sigma_2^2 + 2 \rho \sigma_1 X_t \sigma_2^2}\  dW_t \ .
\end{equation}
This time, from the results of \cite{quantile} we obtain the equilibrium quantile ODE as
\begin{equation}
\frac{\partial^2 Q}{\partial u^2} \biggl(\frac{\partial Q}{\partial u}  \biggr)^{-2} = \frac{2[(\rho \sigma_1 \sigma_2 - \mu_1)+(\sigma_2^2 + \mu_2) Q)]}{(\sigma_1^2+\sigma_2^2 Q^2 + 2\rho \sigma_1 \sigma_2 Q)}\ .\end{equation}
This is then related to the logarithmic derivative of the density function $f(x)$ as 
\begin{equation}
-\frac{1}{f(Q)} \frac{df(Q)}{dQ} = \frac{2[(\rho \sigma_1 \sigma_2 - \mu_1)+(\sigma_2^2 + \mu_2) Q)]}{(\sigma_1^2+\sigma_2^2 Q^2 + 2\rho \sigma_1 \sigma_2 Q)}\ .
\end{equation}
We can solve this ODE and find that, after careful normalization, 
\begin{equation}
f(x) =k \biggl[1+\biggl(\frac{x-\lambda}{a}\biggr)^2 \ \biggr]^{-(\nu+1)/2} \exp \biggl[-\nu_2 \tan^{-1}\biggl(\frac{x-\lambda}{a} \biggr) \biggr]\ , 
\end{equation}
where the parameters are given by
\begin{equation}
\begin{split}
a &= \frac{\sigma_1}{\sigma_2}\sqrt{1-\rho^2}\ ,\\
\lambda &= -\rho\frac{\sigma_1}{\sigma_2}\ ,\\
m &= 1+\frac{\mu_2}{\sigma_2^2}\ ,\\
\nu&= 1+2\frac{\mu_2}{\sigma_2^2}\ ,\\
\nu_2 &= \frac{2(\mu_1\sigma_2+\rho\sigma_1 \mu_2)}{\sigma_1 \sigma_2^2 \sqrt{1-\rho^2}}\ ,\\
k &= \frac{\Gamma\bigl(\frac{\nu+1}{2} \bigr)}{a \sqrt{\pi} \Gamma\bigl(\frac{\nu}{2} \bigr)}
\bigg|\frac{\Gamma\bigl(\frac{\nu+1+i\nu_2}{2} \bigr)}{\Gamma\bigl(\frac{\nu+1}{2} \bigr)}\biggr|^2\ .
\end{split}
\end{equation}
This is the class Pearson Type IV distribution, which is one candidate for a choice of ``skew-Student'' distribution, with a rich variety of skewness and kurtosis in the structure. A useful guide to the properties of the Type IV Pearson is given by Heinrich \cite{heinrich}, who uses the parameter $m$ above, and his $\nu$ is our $\nu_2$, i.e. his density is
\begin{equation}
f(x) =k \biggl[1+\biggl(\frac{x-\lambda}{a}\biggr)^2 \ \biggr]^{-m} \exp \biggl[-\nu \tan^{-1}\biggl(\frac{x-\lambda}{a} \biggr) \biggr]\ .
\end{equation}
Transferring the results of \cite{heinrich} to our own notation\footnote{Users of the standard student `t' are perhaps more used to working with the degrees of freedom parameter $\nu$.} we can identify the mean, provided $\nu>1$, as
\begin{equation}
E[X] =\lambda-\frac{a\nu_2}{\nu-1}\ .
\end{equation}
The variance exists provided $\nu>2$ and is then
\begin{equation}
\Sigma^2 = E[X^2]-E[X]^2 = \frac{a^2((\nu-1)^2+\nu_2^2)}{(\nu-1)^2(\nu-2)}\ .
\end{equation}
The third moment can be calculated provided $\nu>3$ and leads to the normalized skewness as
\begin{equation}
\frac{E[(X-E[X])^3]}{\Sigma^3} = \frac{-4 \nu_2}{\nu-3}\sqrt{\frac{\nu-2}{(\nu-1)^2+\nu_2^2}}\ .
\end{equation}
The fourth moment exists provided $\nu>4$ and may be expressed through {\it excess} kurtosis, which  is
\begin{equation}
\frac{E[(X-E[X])^4]}{\Sigma^4}-3 =\frac{6 \left(\nu _1-3\right) \left(\nu _1-1\right){}^2+6 \left(5 \nu
   _1-11\right) \nu _2^2}{\left(\nu _1-4\right) \left(\nu _1-3\right)
   \left(\left(\nu _1-1\right){}^2+\nu _2^2\right)}\ .\end{equation}
When $\nu_2=0$ the skewness is zero and the excess kurtosis reduces to the well-known expression for the pure Student distribution: $6/(\nu-4)$. We verified the translation of these expressions to our notation by the computation of the moments by simple numerical integration for numerous parameter values.

\section{Towards the full dynamics}
The achievement of an equilibrium is not realistic for most trading periods, especially during a panic of the credit-crunch period. The pure equilibrium analysis above is meant to indicate how a rich variety of distributional types may emerge from a simple model, and no more. We now turn to the full dynamics. We can consider this from both the SDE point of view and in terms of the Fokker-Planck equation for the time-dependent density. 

\subsection{Formal solution of the SDE}
The solution of the SDE can be given explicitly, and gives a structure similar to the functionals of  pairs of L\'{e}vy processes analysed by Carmona, Petit and Yor \cite{carmonayor}, in \cite{barnbook}, where the structure arising from a pair of L\'{e}vy processes is regarded as a generalized OU process. Here we will give the explicit integration of the SDE in the  case of non-zero correlation for the Brownian motions via a correction to the drift somewhat similar to that employed for quanto options. The following version of the argument is based on the idea of treating the SDE in much the same way as one treats the OU SDE, by using an exponential integrating factor. In this case the integrating factor is itself a stochastic quantity.\footnote{I am grateful to D. Crisan and M. Yor for pointing out the method for the zero correlation solution - personal communication, Kyoto 2009.} Starting from
$$
dX_t = (\mu_1-\mu_2 X_t) dt + \sigma_1 dW_{1t}  + \sigma_2 X_t dW_{2t}\ ,
$$
we introduce the stochastic integrating factor
\begin{equation}
I_t = \exp\bigl[-\sigma_2 W_{2t}+ \bigl(\mu_2 + \frac{1}{2}\sigma_2^2 \bigr)t \bigr]
\end{equation}
and observe that
\begin{equation}
dI_t = (\mu_2 + \sigma_2^2)I_t dt - \sigma_2 I_t dW_{2t}\ .
\end{equation}
Now we apply the integrating factor by defining
\begin{equation}
Q_t = X_t I_t
\end{equation}
Then application of Ito's lemma and a short calculation gives
\begin{equation}
dQ_t = I_t ((\mu_1 - \rho \sigma_1 \sigma_2)dt + \sigma_1 dW_{1t})
\end{equation}
and by integration, with $X_0 = 0 = Q_0$, we obtain
\begin{equation}
Q_t = \int_0^t I_s ((\mu_1 - \rho \sigma_1 \sigma_2) ds + \sigma_1 dW_{1s})
\end{equation}
and hence
\begin{equation}
X_t = \int_0^t (I_t^{-1}I_s) ((\mu_1 - \rho \sigma_1 \sigma_2) ds + \sigma_1 dW_{1s})\ .
\end{equation}
Now we observe that
\begin{equation}
I_t^{-1}I_s = \exp\bigl[\sigma_2 (W_{2t}-W_{2s})+ \bigl(\mu_2 + \frac{1}{2}\sigma_2^2 \bigr)(s-t) \bigr]
\end{equation}
and so by a time reversal $u=t-s$, with associated time-reversed variables $\tilde{W}_i$, we can write
\begin{equation}
X_t = \int_0^t ((\mu_1 - \rho \sigma_1 \sigma_2) du + \sigma_1 d\tilde{W}_{1u}) \exp\bigl[\sigma_2 \tilde{W}_{2u} - (\mu_2 + \frac{1}{2}\sigma_2^2)u \bigr]\ .
\end{equation}
This reveals that the solution is the integral of one Brownian motion against the exponential of a second Brownian motion. Thus the process is a clear generalization both of the Asian SDE ($\sigma_1=0$) and the OU SDE ($\sigma_2=0$), with the resulting complications. We can further reduce the expression in terms of the degrees of freedom variable $\nu$, to 
\begin{equation}
X_t = \int_0^t ((\mu_1 - \rho \sigma_1 \sigma_2) du + \sigma_1 d\tilde{W}_{1u}) \exp\bigl[\sigma_2 \tilde{W}_{2u} - \frac{\nu}{2}\sigma_2^2 u \bigr]\ .
\end{equation}
For future reference we note that in the case $\nu=0=\rho=\mu_1$, we have
\begin{equation}
X_t = \sigma_1 \int_0^t d\tilde{W}_{1u} \exp\bigl[\sigma_2 \tilde{W}_{2u}\bigr]\ .
\end{equation}
so that $X_t$ is just the integral of one pure Brownian motion against the exponential of another. By comparison with the hyperbolic OU SDE obtained in Section 5, we can observe the recovery of Bougerol's identity \cite{bougerol,yorbook}, that in this case $X_t$ is distributionally the $\sinh$ of another Brownian motion. We shall see this special case emerge again when we solve the Fokker-Planck equation.
\subsection{Dynamic moment evolution}
A partial characterization of the full dynamical situation may be given in terms of the evolution of the moments. Let us set
\begin{equation}
e_n = E[X_t^n]\ .
\end{equation}
Then elementary analysis gives us the sequential families of ODEs:
\begin{equation}
\frac{de_n}{dt} + (\mu_2 n - \frac{1}{2}n(n-1)\sigma_2^2) e_n =  \frac{1}{2}n(n-1)\sigma_1^2 e_{n-2} + (\mu_1 n + n(n-1)\rho \sigma_1 \sigma_2)e_{n-1}\ ,
\end{equation}
with $e_0 \equiv 1$. 
The evolution of the mean $e_1$ is therefore governed by 
\begin{equation}
\frac{de_1}{dt} + \mu_2 e_1 =  \mu_1
\end{equation}
and, as in an ordinary OU process, evolves according to
\begin{equation}
e_1 = X_0 e^{-\mu_2 t} + \frac{\mu_1}{\mu_2}(1 - e^{-\mu_2 t})\ .
\end{equation}
With our conventions $X_0 = 0$ at the start of the trading period so we then have
\begin{equation}
e_1 =  \frac{\mu_1}{\mu_2}(1 - e^{-\mu_2 t})\ ,
\end{equation}
which will settle down to $\mu_1/\mu_2$ if $\mu_2>0$ and grows exponentially otherwise.  

\subsection{The explosion of variance}
Solution for the higher moments is straightforward but leads to rather unwieldy formulae in general. In the special case where $\rho=0=\mu_1$ the variance $V(X_t)$ may be written in the tractable form
\begin{equation}
V(X_t) = \frac{\sigma_1^2}{\sigma_2^2 (\nu-2)}\biggl[1- e^{-\sigma_2^2 (\nu-2) t} \biggr] \sim \sigma_1^2 t - \frac{1}{2} \sigma_1^2 \sigma_2^2 (\nu-2)t^2 + O(t^3)\ ,
\end{equation}
where, as before $\nu = 1+2\mu_2/\sigma_2^2$. Once the market has kicked off (the behaviour near $t=0$ being always Gaussian), the market dynamics are thus critically dependent on the sign of $\nu-2$. If the strength of the mean-reverting trades is such that $\mu_2 > \sigma_2^2/2$ the market settles down. If instead $\mu_2 < \sigma_2^2/2$ the variance grows exponentially.  Note that there is a region $0<\mu_2 < \sigma_2^2/2$ where the average level stabilizes but the variance does not. If $\mu_2<0$ both the average level and variance grow exponentially. If $\nu>2$ the distribution settles down to the Student equilibrium already analyzed, the with the condition $\nu>2$ guaranteeing a finite variance. 

The exponential growth in variance is not a new idea - it has been present in the {\it price evolution} model of geometric Brownian motion for decades. What is being suggested here is that technical trade effects in the evolution of the {\it log}-return lead to a variance explosion in the log-returns due to these being a hybrid arithmetic-geometric hybrid. 

We can now return to the famous $25\sigma$ issue. If one's perception is that normal Gaussian behaviour is to be expected, with the classical variance $\sigma_1^2 t$ in the log-returns, then the inclusion of technical trading effects via the hybrid model causes the actual variance to differ by a ratio, that we call the {\it variance explosion factor}
\begin{equation}
V_E(t) = \frac{V(t)}{\sigma_1^2 t}\ .
\end{equation}
In general the variance explosion factor may be found by solving the ODE for $e_2$. In the special case considered above we have
\begin{equation}
V_E(t) = \frac{1}{\sigma_2^2 t (\nu-2)}\biggl[1- e^{-\sigma_2^2 (\nu-2) t} \biggr]\ .
\end{equation}
If the markets are minded to settle to equilibrium this ratio tends to zero. But if $\nu < 2$ we have instead
\begin{equation}
V_E(t) = \frac{1}{\alpha t}\biggl[e^{(\alpha t)}-1 \biggr]\ ;\ \ \ \  \alpha = \sigma_2^2 (2-\nu) > 0 \ .
\end{equation}
To return to the likelihood of $25\sigma$ events, we note that if
\begin{equation}
\alpha \geq 6.4746\ ,\ {\text i.e.}\  \sigma_2^2 (2-\nu) > 6.4746\ ,
\end{equation}
then $V_E(t) > 100$ and a $25\sigma$ event based on an initial perception of variance $\sigma_1^2 t$ is no less likely than a $2.5 \tilde{\sigma}$ event with the right variance (and with a different, non-Gaussian, distribution). Such events may then occur repeatedly if the technical market effects are strong enough, even without incorporating the additional effect of net price pressure due to fundamental trades ($\mu_1 \neq 0$).

We see that a variety of different dynamics are possible, with quite small shifts in the mean-reversion strength of technical trades making a dramatic difference to the return distribution.

\subsection{Dynamic distributional aspects}
Having identified the impact of the arithmetic-geometric hybrid on the moments we now need to understand the full shape of the distribution. If the distribution is significantly fat tailed then this can also amplify the likelihood of extreme movements. There are three ways of proceeding. One approach is to transform the problem to the form analyzed by Wong \cite{wong}. The second is to seek a fresh approach to the Fokker-Planck equation. The third is to exploit the representation of Section 7.1 to obtain further insight from a conditionally Gaussian representation. The representation given by Wong involves the integrals of hypergeometric functions, so further insight is helpful, both to understand the theoretical behaviour and to provide more practical tools for practitioners. 

First we should note that the problem of finding the time-dependent PDF in the case $\rho=0=\mu_1$ can be written down by transforming the problem to that analyzed by Wong \cite{wong}. This is the form also used by Nagahara \cite{nagahara}. In our notation, Wong's solution has $\rho=0=\mu_1$, $\nu=2\alpha$, $\sigma_1=\sigma_2 = \sqrt{2}$. The Wong paper from 1963 is available on-line from E. Wong's web page at \hfill\break

\vspace{-0.1in}
\noindent
\url{http://www.eecs.berkeley.edu/~wong/wong_pubs/wong10.pdf}
\smallskip

An issue for financial analysis is the detailed practical implementation of this density and the extraction of insight. Wong also notes that the solution becomes much more straightforward in the case when $\alpha$ is a positive integer, and gives Gaussian functionals for those cases. In general however, the density is an integral of a hypergeometric function of complex parameters. Such expressions are tractable in an advanced computing environment such as {\it Mathematica}, but present challenges in C or other less mathematical languages, in much the same way as for Asian options. Our approach here will be to obtain some insight beyond the work of Nagahara and Wong by direct analysis of the Fokker-Planck equation and relating the outcomes to the theory of integrals of exponentials of Brownian motion, and to the Legendre equation common in physics. We will see that matters are in fact more straightforward when $\alpha = \nu/2$ is any integer, and this is a manifestation of the discrete symmetry of the Legendre equation. Negative cases correspond to momentum-dominated markets, and positive cases are mean-reverting and asymptotically Student. 

A further point of practical implementation of the Wong formula is that given in general it involves complex integrals of hypergeometric functions, it is not {\it necessarily} of greater ease of use than inversion of the transform that we shall derive here, which is also hypergeometric. The underlying issue is whether it is helpful to separate out the discrete eigenvalues of the Sturm-Liouville operator from the continuous ones. The Wong formula has this separation - the transform representation here does not. Both require an inversion by complex integration in general. The derivation from first principles of the transform given here al least offers an alternative, and indeed we can transform it into different forms to make the analysis of positive and negative $\nu$ more transparent. It is hoped that the options for representation here will stimulate further analysis. The matter is somewhat similar to the older analytical studies of Asian options triggered by the Geman-Yor model \cite{gyasian}. Direct inversion of the transform in {\it Mathematica} was possible with a few lines of code \cite{shawbook}, as long as the parameter $\sigma^2 T$ was not too small. The work by Linetsky \cite{linetsky} established the useful spectral representation, but probably the simplest solution to the low volatility evaluation problem is an asymptotic analysis \cite{ejam,zhangb}.

To get at the PDF in general we must analyze the full Fokker-Planck equation in the form

\begin{equation}
 \frac{\partial f(x,t)}{\partial t} =  \frac{\partial \ }{\partial x} \biggl[ -(\mu_1-\mu_2 x) f(x,t) + \frac{1}{2}\frac{\partial  }{\partial x}\bigl[ (\sigma_1^2 +x^2 \sigma_2^2 + 2 \rho \sigma_1 \sigma_2 x) f(x,t))\bigr]  \biggr]\ ,
\end{equation}
with the initial condition $f(x,0)=\delta(x)$. One route to this is via the Laplace transform with respect to time. So let
\begin{equation}
\tilde{f}(x,p) = \int_0^{\infty}f(x,t) e^{-p t}dt\ .
\end{equation}
Then the Fokker-Planck equation gives us, suppressing the independent variables, 

\begin{equation}
p \tilde{f} - \delta(x,0) =  \frac{\partial \ }{\partial x} \biggl[ -(\mu_1-\mu_2 x)  \tilde{f} + \frac{1}{2}\frac{\partial  }{\partial x}\bigl[ (\sigma_1^2 +x^2 \sigma_2^2 + 2 \rho \sigma_1 \sigma_2 x) \tilde{f})\bigr]  \biggr]\ ,
\end{equation}
This is now a Green's function computation on the transform. For $x > 0$ and $x<0$ we need two independent solutions of 
\begin{equation}
p \tilde{f}  =  \frac{\partial \ }{\partial x} \biggl[ -(\mu_1-\mu_2 x)  \tilde{f} + \frac{1}{2}\frac{\partial  }{\partial x}\bigl[ (\sigma_1^2 +x^2 \sigma_2^2 + 2 \rho \sigma_1 \sigma_2 x) \tilde{f})\bigr]  \biggr]\ ,
\end{equation}
with the junction condition that $\tilde{f}$ is continuous at $x=0$, and a jump in the first derivative. This condition, integrating about zero, is
\begin{equation}
\frac{\partial \tilde{f}}{\partial x}(0+, p)- \frac{\partial \tilde{f}}{\partial x}(0-, p)= -\frac{2}{\sigma_1^2}\ .
\end{equation}

\subsubsection{The Gaussian case}
In order to make this approach clearer, we first pursue the standard Gaussian problem where $\sigma_2 = 0 = \mu_2$. The solution of the transformed ODE vanishing as $x \rightarrow \pm \infty$ and with the correct junction condition at zero is
\begin{equation}
\tilde{f}(x,p) = \frac{e^{\mu_1 x/\sigma_1^2}}{\sqrt{\mu_1^2+2 p \sigma_1^2}}  
\begin{cases} \exp\bigl[-\frac{x}{\sigma_1^2}\sqrt{\mu_1^2 + 2 p \sigma_1^2}  \bigr] & \text{if $x>0$,}
\\
\exp\bigl[+\frac{x}{\sigma_1^2}\sqrt{\mu_1^2 + 2 p \sigma_1^2}  \bigr]  &\text{if $x<  0$.}
\end{cases}
\end{equation}
Inversion of the two cases leads to the single well-known formula
\begin{equation}
f(x, t) = \frac{1}{\sqrt{2 \pi \sigma_1^2 t}} \exp\bigl[{-(x-\mu_1 t)^2/(2 \sigma_1^2 t)}\bigr]\ .
\end{equation}

\subsubsection{A dynamic Student distribution}
The next case of interest is when $\rho=0=\mu_1$, for which we previously demonstrated a Student t equilibrium under certain circumstances. The Laplace transform of the Fokker-Planck equation is, for $x \neq 0$,
\begin{equation}
\frac{1}{2}(\sigma_1^2 + \sigma_2^2 x^2) \tilde{f}''(x, p) + (\mu_2 + 2\sigma_2^2)x \tilde{f}'(x, p) + (\mu_2 + \sigma_2^2 - p) \tilde{f}(x, p) =0\ .\label{basicfp}
\end{equation}
This equation may be simplified somewhat by setting
\begin{equation}
\tilde{f}(x,p) = (\sigma_1^2 + \sigma_2^2 x^2)^{-(1+\mu_2/\sigma_2^2)}g(x,p)\ ,
\end{equation}
and the ODE for $g(x,p)$ is then
\begin{equation}
(\sigma_1^2 + \sigma_2^2 x^2) g''(x, p) - 2 x \mu_2  g'(x, p) - 2p g(x,p) =0\ .
\end{equation}
We have already worked out the equilibrium case when $p=0$ and $g$ is constant in $x$.  The management of such an equation is straightforward in the special case $\mu_2 = -\sigma_2^2/2$, as discussed in \cite{polyanin}. We shall use the change of independent variables indicated in \cite{polyanin} to treat the general case, and indeed this is almost the same change of variables that took us to the hyperbolic OU picture. We introduce $z(x)$ with the condition that
\begin{equation}
\frac{dz}{dx} = \frac{1}{\sqrt{\sigma_1^2 + \sigma_2^2 x^2}} 
\end{equation}
and fix the arbitrary constants so that
\begin{equation}
z = \frac{1}{\sigma_2} \sinh^{-1}\biggl(\frac{\sigma_2 x}{\sigma_1} \biggr)\ . \label{start}
\end{equation}
Our equation for $g$ expressed in terms of $z$ is then just
\begin{equation}
 \frac{d^2 g}{dz^2} - (2 \mu_2+\sigma_2^2)\frac{1}{\sigma_2} \tanh(\sigma_2 z)  \frac{dg}{dz} -2p g =0\ ,
\end{equation}
or in terms of the degrees of freedom parameter $\nu = 1+2\mu_2/\sigma_2^2$, 
\begin{equation}
 \frac{d^2 g}{dz^2} - \nu \sigma_2 \tanh(\sigma_2 z)  \frac{dg}{dz} -2p g =0\ .   \label{finish}
\end{equation}
We previously remarked on the Bougerol identity \cite{bougerol} for the case when $\nu=0$. This finds a simple expression here (see also Section 3 of Matsumoto and Yor\cite{matsumoto}). In this case the transform is 
\begin{equation}
\tilde{f}(x,p) = \frac{1}{\sqrt{\sigma_1^2 +\sigma_2^2 x^2}}   \frac{1}{\sqrt
{2p}}  
\begin{cases} e^{-\sqrt{2p} z(x)}& \text{if $z, x>0$,}
\\
e^{+\sqrt{2p} z(x)} &\text{if $z, x<  0$.}
\end{cases}\label{simplap}
\end{equation}
and the inversion gives us
\begin{equation}
f(x, t) = \frac{1}{\sqrt{2 \pi t (\sigma_1^2 + \sigma_2^2 x^2)}} \exp\biggl\{\frac{-1}{2 \sigma_2^2 t} [\sinh^{-1}(\sigma_2 x/\sigma_1)]^2 \ ,\biggr\}
\end{equation}
which is the density arising from a change of variables $z \rightarrow x$ on the $z$-density 
\begin{equation}
\frac{1}{\sqrt{2\pi t}} \exp(-z^2/(2 t))\ ,
\end{equation}
and is a significantly fatter-tailed object (while in the neighbourhood of $x=0$ resembling a Gaussian with variance $\sigma_1^2 t$!)
The presence of significant momentum trading has fattened the tails in this case. We shall now explore the properties of the PDF given for the special case hybrid given by Equation (76).

\subsection{The hidden menace}

\begin{figure}[hbt]
\begin{center}
\includegraphics[scale=1]{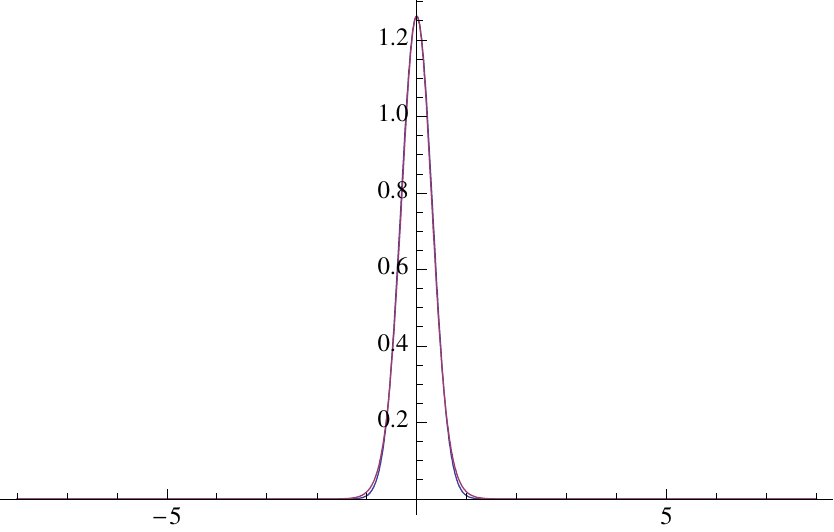}
\end{center}
\caption{PDFs for Gaussian and special case hybrid, $t=0.1$.}\label{grapha}
\end{figure}

\begin{figure}[hbt]
\begin{center}
\includegraphics[scale=1]{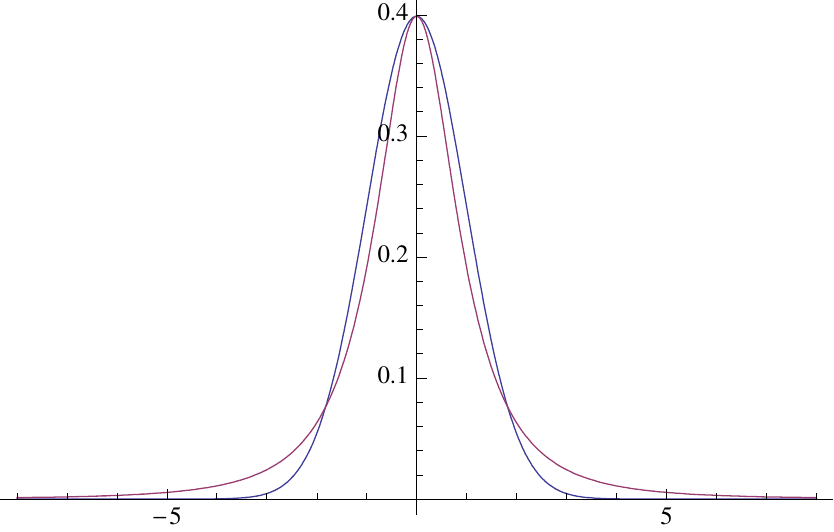}
\end{center}
\caption{PDFs for Gaussian and special case hybrid, $t=1$.}\label{graphb}
\end{figure}

\begin{figure}[hbt]
\begin{center}
\includegraphics[scale=1]{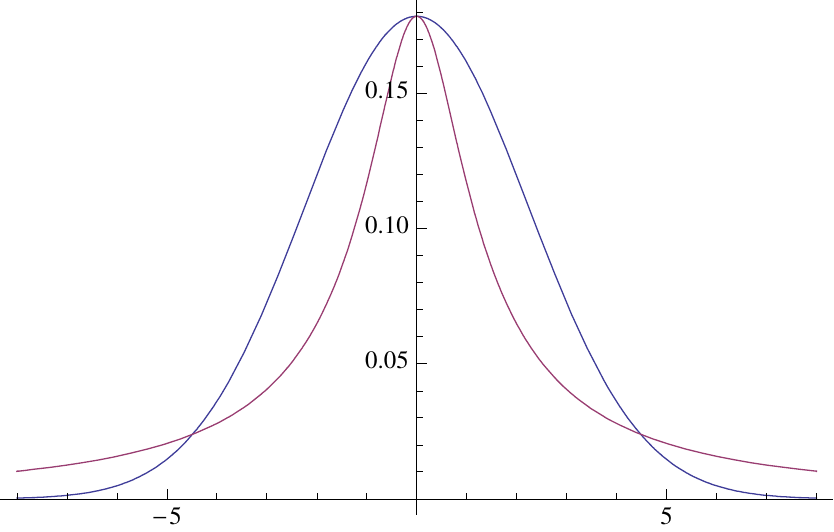}
\end{center}
\caption{PDFs for Gaussian and special case hybrid, $t=5$.}\label{graphc}
\end{figure}

Even the very special case solved for exhibits some interesting features. At the start of the trading period the distribution of log-returns is barely distinguishable from Gaussian, as shown in Figure~\ref{grapha} for the parameters $\sigma_1 = \sigma_2 = 1, t=0.1$, where the hybrid is overlaid with the Gaussian.

However, as time passes the hybrid distribution spreads out more, in a manner consistent with the variance explosion formula. In Figure~\ref{graphb} and Figure~\ref{graphc} we show the hybrid PDF overlaid with the Gaussian at times $t=1$ and $t=5$ respectively. At {\it all} times the probability of a very small move remains at the Gaussian level - the PDF osculates the Gaussian at the origin. The overall behaviour represents the hidden menace of these processes. It starts off looking Gaussian with variance $\sigma_1^2 t$; the probability of a very small movement remains near the Gaussian value, yet dependent on the size of $\sigma_2$ the probability of extreme movements grows exponentially in time.

\section{More general solutions}
We now look at the case where $\nu\geq 0$. We go back to Eqns.~(\ref{start}-\ref{finish}), and change independent variable to
\begin{equation}
u = \sinh^{-1}(\sigma_2 x/\sigma_1) = \sigma_2 z\ .
\end{equation}
Setting $s = 2p/\sigma_2^2$, and letting $'$ now denote $d/du$, Eqn.~(\ref{finish}) may be reorganized as
\begin{equation}
(e^u + e^{-u})(g''(u) - s g(u)) = \nu(e^u - e^{-u}) g'(u)\ . \label{hypode}
\end{equation}
Previously we considered the case $\nu=0$, where we could write down a solution decaying as $x \rightarrow +\infty$ as a single exponential in $e^{-\sqrt{s}u}$.
When $\nu \neq 0$ we have to proceed differently, but the presence of the two exponentials in the coefficients suggests a solution approach. To tidy up a little we make a further change of variable to
\begin{equation}
w = e^{-u} = \biggl(\frac{\sigma_2 x}{\sigma_1} + \sqrt{1 + \frac{\sigma_2^2 x^2}{\sigma_1^2}}\biggr)^{-1}\ .
\end{equation}
So with $u = -\log w$ we have $d/du = -w d/dw$ and our ODE may be rewritten as
\begin{equation}
\biggl(\frac{1}{w} + w\biggr)\biggl[w^2 \frac{d^2 g}{d w^2} + w \frac{d g}{d w} - sg\biggr] = \nu \biggl(w-\frac{1}{w} \biggr)w \frac{d g}{d w}\ .
\end{equation}
We seek a power series solution in the form
\begin{equation}
g = \sum_{k=0}^{\infty} a_k w^{k+\gamma}\ ,\ \ a_0 \neq 0\ .
\end{equation}
After some standard manipulations we find an indicial equation in the form
\begin{equation}
\gamma^2 + \nu \gamma-s = 0\ ,
\end{equation}
and the required root, for $\nu\geq0$, to get the right behaviour  as $x \rightarrow \infty$, $w \rightarrow 0$, is
\begin{equation}
\gamma = \sqrt{s+\frac{\nu^2}{4}}-\frac{\nu}{2}\ .
\end{equation}
The resulting recurrence relation simplifies to
\begin{equation}
a_{k+2}(k+2)(k+2+2\gamma + \nu) = -a_k(k-\nu)(k+2\gamma)\ \ .
\end{equation}
After some experimentation with hypergeometric functions and some work with {\it Mathematica} we are able to recognize the solution in the form
\begin{equation}
g = a_0(p) w ^{\gamma } \, _2F_1\left(\gamma ,-\frac{\nu }{2};\gamma +\frac{\nu
   }{2}+1;-w ^2\right)\ .
\end{equation}
This representation of the solution has the nice property that we can see that the hypergeometric function reduces to a polynomial if $\nu$ is an even integer. The case $\nu=0$ has already been exhibited. Before discussing other such simple cases we must complete the solution and determine $a_0$.  This involves the application of the jump condition on the derivative at the origin, assuming an even solution. After some algebra we find that
\begin{equation}
a_0(p) = \frac{\sigma_1^{\nu}}{\sigma_2 \Omega(\nu, \gamma)}\ ,
\end{equation}
where 
\begin{equation}
\Omega(\nu, \gamma) = \frac{d\ }{dw}\biggl[w ^{\gamma } \, _2F_1\left(\gamma ,-\frac{\nu }{2};\gamma +\frac{\nu
   }{2}+1;-w^2\right)\biggr]\bigg|_{w=1}\ .
\end{equation}
After some use of {\it Kummer's identity} and variations (specifically identities 15.1.21 and 15.1.22 from \cite{amsteg}), we are lead to
\begin{equation}
\Omega(\nu, \gamma) = \frac{2^{1-\gamma } \sqrt{\pi } \Gamma \left(\gamma +\frac{\nu
   }{2}+1\right)}{\Gamma \left(\frac{\gamma }{2}\right) \Gamma
   \left(\frac{1}{2} (\gamma +\nu +1)\right)}\ .
\end{equation}
We arrive at a closed form for the Laplace transform of the density as
\begin{equation}
\tilde{f}(x,p) = \frac{\sigma _1^{\nu } 2^{\gamma -1} w ^{\gamma } \Gamma \left(\frac{\gamma
   }{2}\right) \Gamma \left(\frac{1}{2} (\gamma +\nu +1)\right) \,
   _2F_1\left(\gamma ,-\frac{\nu }{2};\gamma +\frac{\nu }{2}+1;-w
   ^2\right) }{\sqrt{\pi } \sigma _2 \Gamma \left(\gamma
   +\frac{\nu }{2}+1\right)\left(\sigma _1^2+x^2 \sigma
   _2^2\right){}^{\frac{1}{2} (\nu+1)}}\ .\label{denslt}
\end{equation}
We remind the reader that in the use of this expression,
\begin{equation}
\begin{split}
s &= \frac{2p}{\sigma_2^2}\ ,\\
\gamma &= \sqrt{s+\frac{\nu^2}{4}}-\frac{\nu}{2}\ ,\\
w & =  \biggl(\frac{\sigma_2 x}{\sigma_1} + \sqrt{1 + \frac{\sigma_2^2 x^2}{\sigma_1^2}}\biggr)^{-1}\ .
\end{split}
\end{equation}
It is now easy to check the known special case, when $\nu=0$, for the hypergeometric function simplifies leading to
\begin{equation}
\tilde{f}(x,p) =\frac{w ^{\gamma }}{\gamma  \sigma _2 \sqrt{\sigma _1^2+x^2 \sigma
   _2^2}}\ ,\ \ \gamma = \sqrt{s}\ .
\end{equation}
We now also have a new family of relatively simple cases when $\nu$ is an even integer. For example, when $\nu=2$ we have
\begin{equation}
\tilde{f}(x,p) =\frac{\sigma _1^2 }{2  \sigma _2 \left(\sigma
   _1^2+x^2 \sigma _2^2\right){}^{3/2}}\biggl(\frac{w^{\gamma}}{\gamma}+ \frac{w^{\gamma+2}}{\gamma+2} \biggr)\ ,\ \ \gamma = \sqrt{s+1}-1\ .
\end{equation}
In the case $\nu=4$ we have
\begin{equation}
\begin{split}
\tilde{f}(x,p) &=\frac{\sigma _1^4 }{4  \sigma _2 \left(\sigma
   _1^2+x^2 \sigma _2^2\right){}^{5/2}}\biggl(\frac{(3+\gamma)w^{\gamma}}{\gamma(2+\gamma)}+ \frac{2w^{\gamma+2}}{\gamma+2} + \frac{(\gamma+1)w^{\gamma+4}}{(\gamma+2)(\gamma+4)}\biggr)\ ,\\
   \gamma &= \sqrt{s+4}-2\ .
\end{split}
\end{equation}
\subsection{A Legendre representation}
The solution for the Laplace transform when $\mu_1=\rho=0$ has now been characterized. However, the representation above is not the only one that can be given. The solution is in the form of a hypergeometric function $_2F_1(a,b,a-b+1,-w^2)$, where $a = \gamma, b=-\nu/2$ in this case. Such a pattern of arguments to $_2F_1$ leads to several other equivalent representations. For example, five others are given in \cite{amsteg}. One of these is of particular note, and is associated with identity 15.4.15 of \cite{amsteg}. This is the relation, valid for $-\infty < z < 0$,
\begin{equation}
_2F_1(a,b,a-b+1,z) = \Gamma(a-b+1)(1-z)^{-b}(-z)^{b/2-a/2}P_{-b}^{b-a}\biggl(\frac{1+z}{1-z} \biggr)\ ,
\end{equation}
where $P_L^M$ is the associated Legendre function. Making the relevant substitutions with $z = -w^2$ and simplifying the result gives us another formula for the transform:
\begin{equation}
\tilde{f}(x,p) = \frac{2^{\gamma-1+\nu/2}}{\sqrt{\pi}\sigma_1 \sigma_2}\Gamma\bigl(\frac{\gamma}{2}\bigr)\Gamma\bigl(\frac{\gamma+\nu+1}{2}\bigr)(\cosh u)^{-(\nu/2+1)}P_{\nu/2}^{-\nu/2-\gamma}(|\tanh u|)\ ,  \label{legtfm}
\end{equation}
where $u= \sinh^{-1}(\sigma_2 x/\sigma_1)$. This gives us a nice interpretation of the result. The quantity $L = \nu/2$ is the quantity in  physics normally associated with the angular momentum of a quantum system such as the hydrogen atom. The Legendre functions in that case arise naturally via separation of variables of the Laplacian operator. The case $L$ an integer is particularly simple as then the associated Legendre functions can be written in terms of polynomials. While these are not the ordinary Legendre polynomials (the quantity $M$ here is a complex transform variable and not an integer) they are well known - see e.g. \cite{wolfleg} for explicit forms for the first ten. 
Armed with the hindsight of Eqn.~(\ref{legtfm}) one can now revisit the entire solution process and make a change of variables to reduce the transformed Fokker-Planck equation to Legendre's equation. By making the change of variables
\begin{equation}
\tilde{f}(x,p) = \biggl(\frac{1}{\sqrt{1+y^2}} \biggr)^{\nu/2+1} h \biggl(\frac{y}{\sqrt{1+y^2}}  \biggr)\ ,\ \ y = \frac{\sigma_2 x}{\sigma_1}
\end{equation}
the transformed Fokker-Planck equation can be written as
\begin{equation}
\frac{d\ }{dq}\bigl[(1-q^2) \frac{dh}{dq}\bigr] + \biggl[\frac{\nu}{2}\bigl(\frac{\nu}{2}+1\bigr) - \frac{2p/\sigma_2^2+\nu^2/4}{(1-q^2)} \biggr] h(q)=0
\end{equation}
which is the Legendre equation with parameters $L=\nu/2$ and $M^2 = 2p/\sigma_2^2 + \nu^2/4$. 
Such equivalent expressions reveal to us that the transformed distribution is a well-known mathematical object, and perhaps allow evaluation in computation systems where a full implementation of $_2F_1$ might not be available. However, for further analysis here we will work with the original hypergeometric system and the Legendre representation. The hypergeometric form is well adapted to an analysis of the tails, due to the power behaviour in $w$, and there are hints that this variable may be more useful in the full general case where $\mu_1 \neq 0$. There are a limited number of known inversions of Legendre functions where the transform variable is an index to the function and these are also under investigation. For now we turn to another special case for more detailed study. 
\subsection{A ``chameleon'' distribution}
In this sub-section we shall introduce a one-parameter family of dynamic distributions with the following interesting properties:
\begin{itemize}
\item The distributions arise from stochastic differential equations;
\item The mean is identically zero;
\item The variance is of the form $\sigma_1^2 t$;
\item The behaviour is initially Gaussian, in standard form.
\item The distribution tends to a non-Gaussian steady-state.
\end{itemize}
This is just a matter of specializing the analysis above to the case $\nu=2$. Writing out the solution for the transform more explicitly, we have
\begin{equation}
f(x,p) = \frac{\sigma_1^2}{2 \sigma_2 (\sigma_1^2 + \sigma_2^2 x^2)^{3/2}}\biggl[\frac{e^{-[\sqrt{s+1}-1]|u(x)|]}}{\sqrt{s+1}-1} +\frac{e^{-[\sqrt{s+1}+1]|u(x)|]}}{\sqrt{s+1}+1} \biggr]\ ,
\end{equation}
where
\begin{equation}
s = \frac{2p}{\sigma_2^2}\ ,\ \ u(x) = \sinh^{-1}(\sigma_2 x/\sigma_1)\ .
\end{equation}
This may be inverted in closed form, making careful use of identity 29.3.88 from \cite{amsteg} and some standard Laplace transform identities. After some careful simplifications we are lead to the following density function:
\begin{equation}
\begin{split}
f(x,t) &= \frac{ \sigma _1 \exp[-\frac{u(x)^2}{2 t \sigma
   _2^2}-\frac{t \sigma _2^2}{2}]}{\sqrt{2 \pi t
   } \left(\sigma _1^2+x^2 \sigma _2^2\right)}\\
&+\frac{\sigma _2 \sigma _1^2}{2 \left(\sigma _1^2+x^2 \sigma
   _2^2\right){}^{3/2}}\biggl[    \Phi
   \left(\frac{|u(x)|+t \sigma _2^2}{\sqrt{t} \sigma
   _2}\right)-   \Phi \left(\frac{|u(x)|-t \sigma
   _2^2}{\sqrt{t} \sigma _2}\right)  \biggr] \ ,
\end{split}   
\end{equation}
where $\Phi$ is the standard normal CDF. This is the probability density function for our ``chameleon distribution''. We have obtained it by solving the Fokker-Planck equation from an SDE. Its mean is zero and is variance satisfies
\begin{equation}
V(X) = \sigma_1^2 t \ ,\ \ \   \forall \sigma_2,\  t\ .
\end{equation}
Its asymptotic behaviour as $t \rightarrow 0$ is obtained by considering just the first exponential part of the expression, which we see tends to
\begin{equation}
f(x,t) \sim \frac{ \sigma _1 \exp[-\frac{u(x)^2}{2 t \sigma_2^2}]}{\sqrt{2 \pi t
   } \left(\sigma _1^2+x^2 \sigma _2^2\right)}  \sim \frac{\exp[-\frac{x^2}{2 t \sigma_1^2}]}{\sqrt{2 \pi t
   } \sigma _1}\ ,
   \end{equation}
   where the last approximation arises as the support of the distribution contracts about the origin, allowing us to expand the arcsinh and denominator. So the distribution starts off in standard Gaussian form. For $t \rightarrow \infty$ we just note that the first term tends to zero and the second line tends to
   \begin{equation}
f(x,t) \sim \frac{\sigma _2 \sigma _1^2}{2 \left(\sigma _1^2+x^2 \sigma
   _2^2\right){}^{3/2}}\ ,
\end{equation}
which is the density of a scaled Student t distribution with two degrees of freedom. This of course has infinite variance. We have therefore demonstrated the list of conditions claimed at the start of this Section. Parametrized by $\sigma_2$, there are infinitely many SDEs of the form
\begin{equation}
dX_t = -\frac{\sigma_2^2}{2} X_t dt + \sqrt{\sigma_1^2 + \sigma_2^2 X_t^2} dW_t \ , \label{chamsde}
\end{equation}
whose variance is given by the standard formula $\sigma_1^2 t$, normally associated with the simple case
\begin{equation}
dX_t =  \sigma_1 dW_t\ .
\end{equation}
This is a useful reminder that {\it linear evolution of variance does not in any way imply the elementary Brownian model}. Although we have produced a rather exotic density function, the underlying dynamics as given by Eqn.~(\ref{chamsde}) are very simple. 

\subsection{Momentum-dominated markets}
We now consider the case $\nu < 0$. We can go back to the Fokker-Planck equation~(\ref{basicfp}) and seek more appropriate changes of variables, or explore the continuation of the $\nu \geq 0$ procedure to $\nu < 0$, and observe that it must remain valid. Equivalently we can exploit the symmetry of the Legendre representation under $L \rightarrow -(L+1)$ to derive the appropriate transform. Perhaps the simplest approach is to exploit a hypergeometric identity (see e.g. \cite{amsteg}), that is also an encoding of the Legendre symmetry:
\begin{equation}
_2F_1(c-a,c-b,c,z)= (1-z)^{(a+b-c)} {}_2F_1 (a,b,c,z)
\end{equation}
We apply this identity with $a = \gamma, b = -\nu/2, c = \gamma+\nu/2+1, z= -w^2$
\begin{equation}
_2F_1(\gamma,\frac{\nu}{2},\gamma+\frac{\nu}{2}+1,-w^2)= (1+w^2)^{(\nu+1)} {}_2F_1 (\frac{\nu}{2}+1,\gamma+\nu+1,\gamma+\frac{\nu}{2}+1,-w^2)
\end{equation}
Next, given the form of $w$, we note that
\begin{equation}
1+w^2 = \frac{2w}{\sigma_1}\sqrt{\sigma_1^2 + \sigma_2^2 x^2}
\end{equation}
and putting these observations together with the transform density representation Eqn.~(\ref{denslt}) we obtain a representation better adapted to $\nu <0$:
\begin{equation}
\begin{split}
\tilde{f}(x,p) =& \frac{ 2^{\gamma +\nu} w ^{\gamma+\nu+1 } \Gamma \left(\frac{\gamma
   }{2}\right) \Gamma \left(\frac{1}{2} (\gamma +\nu +1)\right) }{\sqrt{\pi } \sigma _1 \sigma _2 \Gamma \left(\gamma
   +\frac{\nu }{2}+1\right)}
   \\
 &  \times
   {}_2F_1(\frac{\nu}{2}+1,\gamma+\nu+1,\gamma+\frac{\nu}{2}+1,-w^2) \ .\label{densltneg}
\end{split}
\end{equation}
where we also recall that
$$
\gamma = \sqrt{s+\frac{\nu^2}{4}}-\frac{\nu}{2}\ .$$
We now see that another set of special and simple cases emerge. For $\nu=-2$ we have, simplifying,
\begin{equation}
\tilde{f}(x,p) = \frac{w^{\gamma-1}}{\sigma_1 \sigma_2 (\gamma-1)}=\frac{w^{\sqrt{s+1}}}{\sigma_1 \sigma_2 \sqrt{s+1}}\ .
\end{equation}
For $\nu=-4$ we have, simplifying,
\begin{equation}
\begin{split}
\tilde{f}(x,p) &= \frac{1}{2\sigma_1 \sigma_2}\biggl(\frac{w^{\gamma-1}}{\gamma-1} +\frac{w^{\gamma-3}}{\gamma-3}\biggr)\\
\gamma &= \sqrt{s+4}+2
\end{split}
\end{equation}
And for $\nu=-6$ we have, simplifying,
\begin{equation}
\begin{split}
\tilde{f}(x,p) &=
\frac{1}{4\sigma_1 \sigma_2}\biggl(\frac{(\gamma -2) w^{\gamma
   -5}}{(\gamma -5) (\gamma
   -3)}+\frac{2 w^{\gamma
   -3}}{\gamma
   -3}+\frac{(\gamma -4)
   w^{\gamma -1}}{(\gamma -3)
   (\gamma -1)}\biggr)\\
   \gamma &= \sqrt{s+9}+3
\end{split}
\end{equation}
These of course mirror the three cases $\nu=0,2,4$ described earlier and exhibit the Legendre symmetry, but without the Student denominators, as is natural for the increasingly explosive market behaviour that these new cases describe.
\subsection{Generalized Bougerol Identity}
The case $\nu=-2$ merits further study as it is the simplest momentum-dominated case, and it turns out links to some nice insights on hyperbolic Brownian motion \cite{revistapaper}, as also summarized in \cite{scandibook}. The $x$-space density is obtained by inversion as
\begin{equation}
f(x,t) = \frac{1}{\sigma_2 \sqrt{2\pi t}} \exp(-\frac{\sigma_2^2 t}{2} - \frac{u^2}{2 \sigma_2^2 t} )
\end{equation}
where 
$$ u = \sinh^{-1}\frac{\sigma_2 x}{\sigma_1}$$
Now transforming to the $u$-density, with $dx \rightarrow \frac{\sigma_1}{\sigma_2}\cosh(u) du$, we obtain the $u$-space density as:
\begin{equation}
g(u,t) = \frac{\sigma_1}{\sigma_2}\cosh(u)f(x,t) = \frac{1}{\sigma_2\sqrt{2\pi t}}\biggl( e^{-(u-\sigma_2^2 t)^2/(2 \sigma_2^2 t)} + e^{-(u+\sigma_2^2 t)^2/(2 \sigma_2^2 t)} \biggr)
\end{equation}
so that in the dimensionless time coordinate $\tau=\sigma_2^2 t$ we see that the variable $\sigma_2 X_t/\sigma_1$ is distributed as
\begin{equation}
\sinh(W_{\tau} + Y \tau)
\end{equation}
where $Y$ is Bernoulli and $W_{\tau}$ is a standard Brownian motion.  This corresponds to exponential drift $+1$ when viewed as the exponential of a Brownian motion. It is possible that similar simple interpretations exists for drift $+2$ ($\nu=-4$) and indeed for the simple cases considered earlier $\nu=2,4$ where the densities also reveal a similar superposition of opposite and multiple drifts.

Perhaps the most significant practical consequence is the emergence of a {\it bi-modal} density, where there are essentially two densities moving apart at a constant speed. In the absence of fundamental drift this makes good sense for a momentum-dominated model, and is expected to be a generic feature for $\nu < 0$.

\subsection{Conditionally Gaussian form for $\rho=0$}
The third route to characterizing the density involves the formal solution of the SDE in the form
\begin{equation}
X_t = \int_0^t (\mu_1 du + \sigma_1 d\tilde{W}_{1u}) \exp\bigl[\sigma_2 \tilde{W}_{2u} - \frac{\nu}{2}\sigma_2^2 u \bigr]\ .
\end{equation}
This is valid for zero correlation between the two Brownian motions. If we condition on the exponential Brownian motion\footnote{I am grateful to D Crisan for this idea.} the conditional distribution of $X_t$ is readily seen to be
Gaussian with mean $m$ and variance $v$, where
\begin{equation}
m = \mu_1 \int_0^t du \exp\bigl[\sigma_2 \tilde{W}_{2u} - \frac{\nu}{2}\sigma_2^2 u \bigr]
\end{equation}

\begin{equation}
v  = \sigma_1^2 \int_0^t \exp\bigl[2\sigma_2 \tilde{W}_{2u} - \nu\sigma_2^2 u \bigr]\ .
\end{equation}
This is an elegant representation that makes it clear that the mean and variance are Asian-like quantities. This also opens up the use of other exact and approximate distributional representations - this will be pursued elsewhere. However, we note that this is a natural outcome if we wish to have a conditional Gaussian model whose total distribution is asymptotically a Student t, in that the conditional variance must asymptotically be inverse Gamma. This condition is satisfied by a conditional variance given by an Asian-like variable, i.e. the integral of the exponential of a Brownian motion. That this is created by a simple price-feedback model we think is interesting. One final point is that this type of representation is frequently used in the probabilistic literature. For example, the description of the Bougerol identity as phrased by Yor \cite{yorbook} is in the equivalent representation of a Brownian motion with an Asian time change, e.g. 
\begin{equation}
\sinh B_t \displaystyle_{\ \ \ =\ \ }^{(LAW)}\gamma_{A_t}
\end{equation}
where $B$ and $\gamma$ are independent Brownian motions and $A_t$ is an Asian form.

\section{States of the market}
Having looked at some typical dynamics, we look at the overall picture. Our model has four parameters, and we now give them names:
\begin{itemize}
\item $\sigma_1$, the fundamental volatility;
\item $\sigma_2$, the technical volatility;
\item $\mu_1$, the fundamental drift;
\item $\mu_2$, the technical drift;
\end{itemize}
While we have not incorporated $\mu_1$ into any detailed analysis thus far, its interpretation is clear. The main influences on the state of the market are the parameters $\mu_2$, $\sigma_1, \sigma_2$. In fact, it is the balance between these parameters that matters. A critical quantity is 
\begin{equation}
\nu = 1 + \frac{2\mu_2}{\sigma_2^2}\ .
\end{equation}
If an equilibrium is achieved, this is the degrees of freedom of the associated Student distribution that results. But now we see that it plays an essential dynamical role:
\begin{itemize}
\item if $\nu < 2$ the variance explodes exponentially;
\item if $\nu=2$ the variance remains in Gaussian form, but any member of the chameleon family may exist;
\item $\nu >2$ the variance tends to a constant.
\end{itemize}
The circumstances under which the distribution attains an equilibrium are subtly different. We know that when $\nu=0,\ m=1/2$ the solution is eternally dynamic and the PDF has been calculated explicitly. When $m>1/2, \nu>0$, the equilibrium solution exists and has a normalizable PDF in Pearson IV form \cite{heinrich}. So there is a range $0<\nu < 2$ where the equilibrium exists but the variance explodes. It would be useful to get a better grip on the dynamic Cauchy case that sites in the middle of this zone with $\nu=1=m$. This case corresponds to $\mu_2=0$. When $\nu <0$, and the market is momentum-dominated, matters are completely different and the distribution may become bimodal, as exemplified by the Alili-Dufresne-Yor generalization of the Bougerol identity for $\nu = -2$ \cite{revistapaper}. 

The volatilities themselves play a different role. Inspection of the hyberbolic OU equation indicates that it is $\sigma_2$ that sets the time-scales, and the ratio $\sigma_2/\sigma_1$ determines the scale on which asset price movements are affected by the price feedback. So in the absence of fundamental drift the market state is best characterized by the triple:
\begin{itemize}
\item $\nu=1 + \frac{2\mu_2}{\sigma_2^2}$ - determines the market condition;
\item $\sigma_2$ defines the time-scale;
\item $\sigma_2/\sigma_1$ defines the asset price scale. 
\end{itemize}
Since all but the first are scaling variables, we can observe that when $\mu_1=0$, the parameter $\nu$ is critical. Historically this has been the elementary `degrees of freedom'' parameter for static Student distributions. It now plays a critical dynamical role. If we set $\tau = \sigma_2^2 t$ the underlying dimensionless SDE is
\begin{equation}
dZ_{\tau} = -\frac{\nu}{2} \tanh(Z_{\tau}) d\tau + dW_{\tau} \ .
\end{equation}
with all other parameters removed by transformation. We have explicit time-domain solutions for the resulting PDF for $\nu=0,2$ and the Laplace transform for other values. 

\section{Initial VaR analysis}
The role and controversy around VaR (Value at Risk) have been discussed extensively elsewhere. There is a fundamental issue of principle about estimating the probability of extreme events. We never know when there is a ``tsunami''-type event that falls outside all of our traditional modelling. However, within any proposed model we should also look at estimating the risk numbers, in order to at least give an improvement based on the analysis we have been able to do.

\subsection{Simple Gaussian VaR}

Consider a percentile point $u_0$ in the unit interval $0<u_0<1$. Let $Q$ denote the standard Gaussian quantile function associated with the CDF $\Phi(x)$. So we have the identity
\begin{equation}
\Phi[Q(u_0)] = u_0 \ .
\end{equation}
In a simple analysis we might take $u_0 = 0.25$, for example. In a standard Gaussian model, in the absence of fundamental drift, the signed Var would just be
\begin{equation}
sVar_G = \sigma_1 \sqrt{t} Q(u_0)
\end{equation}
and the VaR would be the absolute value of this scaled Gaussian quantile. 

\subsection{Simple hyperbolic VaR via Bougerol}
The simplest more general case is that we have previously considered, which is the $\nu=0$ market state. Bougerol's identity gives us the density, and the cumulative distribution function is easily computed from the density function in Section 7.3.2, as
\begin{equation}
F(y) = \Phi \biggl[\frac{1}{\sigma_2 \sqrt{t}} \sinh^{-1}\bigl(\frac{\sigma_2 y}{\sigma_1} \bigr) \biggr]
\end{equation}
The inversion of this relation for the quantile is now trivial, and we find that
\begin{equation}
sVaR_H = y = \frac{\sigma_1}{\sigma_2}\sinh \biggl(\sigma_2 \sqrt{t} Q(u_0)\biggr)
\end{equation}
The value at risk is then seen to be hyperbolically greater than the Gaussian case, and the only coincide in the absence of any technical volatility or for very small times.

\subsection{More general hyperbolic VaR}
It is an interesting question to pursue the VaR estimates within the most general model. The Bougerol identity and its speculative generalization might suggest that there is an approximate approach for the momentum-dominated case by adjusting the simple hyperbolic VaR by a drift correction. In the mean-reverting case one might consider using the stationary asymptotic form. This is particularly simple for the Student case, where there is a lot of developed technology for the T-quantile. There are closed forms for $\nu=1,2,4$ - see \cite{shawjcf06}, and the on-line entry at \cite{wikiquantile}, and series solutions for positive real $\nu$ to any required precision \cite{quantile}. Bailey's elegant generalization of the Box-Muller formula and its polar form \cite{bailey} allows rapid adaptation of many embedded risk management models.  A systematic mathematical treatment of an optimal VaR formula for the general case will be pursued elsewhere.

\section{Conclusions and further work}
We have developed a simple hybrid trading model that mixes both fundamental and technical trades. The general form of the model allows for jumps and Brownian motion together, but the theory here has been more fully developed for the pure Brownian case, where we have obtained several insights into the model proposed by Nagahara \cite{nagahara}. We obtain a hybrid process for the log-return that is a composite of arithmetical and geometric types, and that is capable of exhibiting a variety of behaviours depending on the relative strength of the fundamental and technical components. 

Even when the market settles down we obtain a rich family of distributions including fat-tailed Student and skew-Student models as a special case. It is therefore possible to attribute some of the skewness and fat-tailed behaviour in asset returns to a simple composite model of traders acting on different logic. 

In the fully dynamic context a the likelihood of extreme events is greatly increased by the use of a hybrid arithmetical-geometric process for the log-returns. Such hybrids can exhibit the phenomenon of variance explosion at the same time as hiding their non-Gaussian nature very effectively.  Work on  the full time-dependent form, with all parameters non-zero, is in progress, as is the VaR and related analysis.  

Another thread of research involves the use of multivariate extensions of these models, where the coupled distributions arise through coupled diffusions of Pearson type. In this way multivariate distributions with marginals drawn from the Pearson family may be constructed, as discussed by Shaw and Munir (2009) \cite{warwick}.

The adaptation of the model here to full risk-neutral pricing requires further work. The production of a density for the case where the drift is constant needs to be undertaken. But there are two further issues. First, the model here is regarded as adapted to short-term return modelling and the risk consequences. In reality different trading periods will have different market states. Some days will settle due to dominant mean-reversion, while others will have explosive behaviour. So it is not regarded as appropriate to simply replace the log-normal distribution by the ones implied here - a more subtle approach is needed, though one could consider short-term options within this model. A further difficulty is whether the dynamic hedging process remains viable in the presence of the potentially explosive movements within this model in a momentum-dominated state. While it is mathematically possible to define the hedging process, in reality the explosion of variance could result in rapid movements that render it infeasible. 

Overall we wish to note the dramatic consequences of including a simple linear model of price feedback into a stochastic model. The outcomes of a hybrid of arithmetic and geometric Brownian motion are qualitatively more diverse than the possibilities predicted by either in isolation, and a hybrid offers several features observed in the market. In particular variance explosion is predicted, as are power-law densities under certain conditions. Of course, much more realistic simulations might be done based on observed details of the order-book dynamics, better models of price-impact, and the endless non-Markov ideas of actual technical trading. However, while these might be readily simulated on a computer, the much-simplified approach presented here at least offers the hope of analytical tractability and insight. 

\section*{Acknowledgments}
Dr G. Steinbrecher is responsible for providing information on the existing literature on hybrid processes in theoretical physics. Dr A. Macrina gave many useful comments on earlier versions of this article. I also wish to thank K. Vanguelov, R. Wilson and A. Healey for several useful remarks. I am indebted to M. Yor and D. Crisan for many insights into the stochastic aspects, and to the University of Kyoto for their hospitality during a visit in which some of these ideas were refined.
\

\end{document}